\documentclass[10pt, notitlepage]{article}
\pdfoutput=1

\usepackage[english]{babel}

\usepackage{verbatim}
\usepackage{amssymb}
\usepackage{float}
\usepackage{subfig}
\usepackage{graphicx}
\usepackage[T1]{fontenc}
\usepackage[utf8]{inputenc}
\usepackage{authblk}
\usepackage{amsmath,bm}
\usepackage{amsmath,amssymb}
\usepackage{multicol}
\usepackage{amsmath}
\usepackage[left=2.0cm,right=2.0cm,top=2.0cm,bottom=2.0cm]{geometry}
\renewcommand\footnotemark{}

\usepackage[shortlabels]{enumitem}
\usepackage[utf8]{inputenc}

\usepackage{amsmath,amsthm}
\usepackage{cleveref}
\crefname{axiom}{axiom}{axioms}
\Crefname{axiom}{Axiom}{Axioms}

\newtheorem{lemma}{Statement}

\begin{document}
\title{A note on conformal symmetry}
\author{D.N. Coumbe}
\affil{\small{\emph{The Niels Bohr Institute, Copenhagen University \\Blegdamsvej 17, DK-2100 Copenhagen Ø, Denmark.\\E-mail: daniel.coumbe@nbi.ku.dk}}} 
\date{\small({\today})}
\maketitle

%%%%%%%%%%%%%%%%%%%%%%%%%%%%%%%%%%%%%%%%%%%%%%%%%%%%%%%%%%%%%%%%%%%%%%%%%%%%%%%%                                                                                                               
\begin{abstract}

We investigate a conformal-like transformation for which the spacetime interval is invariant. 

\vspace{0.5cm}
%\noindent \small{PACS numbers: 04.60.Gw, 04.60.Nc}\\

\vspace{0.5cm}
\end{abstract}

%%%%%%%%%%%%%%%%%%%%%%%%%%%%%%%%%%%%%%%                                                                                                                                                
\begin{section}{Introduction}

%%%%%%%%%%%%%%%%%%%%%%%%%%%%%%%%%%%%%%%%%%%%%%%%%%%%%%%%%%%%%%%%%%%                                                                                                               

Symmetry principles have proven remarkably successful in discovering the laws of nature. The power of symmetry was first brilliantly demonstrated by Einstein in 1905. Hidden within Maxwell's equations, Einstein noticed a symmetry and promoted it to a symmetry of spacetime itself, directly leading to the special theory of relativity. Increasing the symmetry of spacetime further, by demanding the laws of physics be invariant under any local change of spacetime coordinates, was even more successful, guiding Einstein to the general theory of relativity~\cite{Gross123}. Can this trend be continued? Can enhancing the symmetry of spacetime even further help reveal new laws of nature? 

Under a plausible set of assumptions, the Coleman-Mandula theorem states that the maximal symmetry spacetime can have is conformal symmetry~\cite{PhysRev.159.1251}. Given the privileged status of conformal symmetry, its implementation may prove fruitful in the search for new fundamental physics. In this work, we explore the effect of endowing spacetime with conformal symmetry. 

Global scale transformations act on spacetime coordinates via $x^{\mu} \rightarrow x^{\mu} \Omega$, such that the metric tensor transforms via $g_{\mu \nu} \rightarrow g_{\mu \nu}\Omega^{2}$, where $\Omega$ has the same value at each spacetime point. However, it is now widely believed that all fundamental symmetries of nature must be local, as exemplified by the successes of gauge theories and general relativity~\cite{Gross123}. Conformal transformations are nothing but local scale transformations. A conformal transformation is defined as a coordinate transformation $x^{\mu} \rightarrow \tilde{x}^{\mu}(x)$ such that the metric tensor transforms according to $g_{\mu \nu} \rightarrow g_{\mu \nu}\Omega^{2}(x)$, where $\Omega(x)$ is a positive dimensionless function of all four spacetime coordinates $x=(x^{0},x^{1},x^{2},x^{3})$.

%Global scale transformations act on spacetime coordinates via $x^{\mu} \rightarrow \tilde{x}^{\mu}(x^{\mu})=x^{\mu} \Omega$, such that the metric tensor transforms via $g_{\mu \nu} \rightarrow \tilde{g}_{\mu \nu}=g_{\mu \nu}\Omega^{2}$, where $\Omega$ has the same value at each spacetime point. However, it is now widely believed that all fundamental symmetries of nature must be local, as exemplified by the successes of gauge theories and general relativity~\cite{Gross123}. Conformal transformations are nothing but localised scale transformations. A conformal transformation is defined as a coordinate transformation $x \rightarrow \tilde{x}(x)$ such that the metric tensor transforms according to $g_{\mu \nu} \rightarrow g_{\mu \nu}\Omega^{2}(x)$, where $\Omega(x)$ is a positive dimensionless function of all four spacetime coordinates $x=(x^{0},x^{1},x^{2},x^{3})$.

In this work we consider the coordinate transformation

%A conformal transformation can be defined as a localised point-dependent coordinate transformation 

\begin{equation}\label{eq1}
x^{\mu} \rightarrow \tilde{x}^{\mu}(x^{\mu})=x^{\mu} \Omega(x^{\mu}),
\end{equation}

\noindent where each coordinate is rescaled by a function of that coordinate only ($x^{\mu}$ represents an individual coordinate), such that the metric tensor transforms according to 

\begin{equation}\label{eq2}                                                                                                                                                                
g_{\mu \nu} \rightarrow g_{\mu \nu}\Omega^{2}(x^{\mu}),
\end{equation}

\noindent where $\Omega(x^{\mu})$ is a positive dimensionless function of the individual spacetime coordinate $x^{\mu}$. In section~\ref{condition} we explore the possible connection between Eqs.~(\ref{eq1}) and~(\ref{eq2}) and a conformal transformation.

The aim of this paper is to determine the only non-trivial factor $\Omega(x^{\mu})$ consistent with a locally invariant spacetime. To achieve this aim we demand that the line element $ds^{2}=g_{\mu\nu} dx^{\mu}dx^{\nu}$ be invariant under the transformation of Eqs.~(\ref{eq1}) and~(\ref{eq2}). 

%%%%%%%%%%

\end{section}

%%%%%%%%%%%%%

\begin{section}{Determining $\Omega(x^{\mu})$}\label{motivation2}

%%%%%%%%%%%%%%%%%%%%%%%%%%%%%%%%%%%%%%%%%%%%%%%%%%%%%%%%%%%%%%%%%%%                                                                                                                    
%%%%%%%%%%%%%%%%%%%%%%%%%%%%%%%%%%%%%%

\begin{subsection}{Invariant line element}\label{int}

\begin{lemma}
The only non-trivial factor $\Omega(x^{\mu})$ for which the line element $ds^{2}=g_{\mu\nu} dx^{\mu}dx^{\nu}$ is invariant under the transformation of Eqs.~(\ref{eq1}) and~(\ref{eq2}) is

%The only non-trivial conformal transformation for which the line element $ds^{2}=g_{\mu\nu} dx^{\mu}dx^{\nu}$ is invariant is

%\begin{equation}\label{eq3}
%x^{\mu} \rightarrow x^{\mu}\Omega(x^{\mu}) = x^{\mu} \sqrt{1+ \left(\frac{l}{x^{\mu}}\right)^{2}}, \qquad g_{\mu \nu} \rightarrow g_{\mu \nu}\left(1+ \left(\frac{l}{x^{\mu}}\right)^{2} \right),
%\end{equation}

\begin{equation}\label{eq3}
\Omega(x^{\mu}) = \sqrt{1+ \left(\frac{l}{x^{\mu}}\right)^{2}},
\end{equation}

\noindent where $l$ is a constant with dimensions of length.
\end{lemma}

\begin{proof}[\rm{\textbf{Proof 1}}]

For the line element $ds^{2}=g_{\mu\nu} dx^{\mu}dx^{\nu}$ to be invariant under Eq.~(\ref{eq2}) the product $dx^{\mu}dx^{\nu}$ must transform according to

\begin{equation}\label{proof2}
dx^{\mu}dx^{\nu} \rightarrow \frac{dx^{\mu}dx^{\nu}}{\Omega(x^{\mu})^{2}} \equiv d\tilde{x}^{\mu}d\tilde{x}^{\nu}.
\end{equation}

\noindent For diagonal components ($\mu=\nu$) of $g_{\mu \nu}$ we therefore require

\begin{equation}\label{proof3}
(d\tilde{x}^{\mu})^{2}=\frac{(dx^{\mu})^{2}}{\Omega(x^{\mu})^{2}}.
\end{equation}

\noindent Taking the square root and substituting $\tilde{x}^{\mu}=x^{\mu} \Omega(x^{\mu})$ into Eq.~(\ref{proof3}) gives

\begin{equation}\label{proof3.5}
\frac{d\left(x^{\mu}\Omega(x^{\mu})\right)}{dx^{\mu}} - \frac{1}{\Omega(x^{\mu})}=0.
\end{equation}

\noindent Since the product rule tells us that

\begin{equation}\label{proof3.6}
\frac{d\left(x^{\mu}\Omega(x^{\mu})\right)}{dx^{\mu}}= \Omega(x^{\mu}) + x^{\mu}\frac{d\Omega(x^{\mu})}{dx^{\mu}},
\end{equation}

\noindent we have

\begin{equation}\label{proof4}
\Omega(x^{\mu})+ x^{\mu} \frac{d\Omega(x^{\mu})}{dx^{\mu}} - \frac{1}{\Omega(x^{\mu})}=0.
\end{equation}

\noindent Equation~(\ref{proof4}) must be satisfied in order to yield an invariant line element. Equation~(\ref{proof4}) has four solutions

\begin{equation}\label{proof5}
\Omega(x^{\mu})= \left\{\pm 1,\; \pm \sqrt{1+ \frac{a}{(x^{\mu})^{2}}}\right\},
\end{equation}

\noindent where $a$ is a constant. Since $\Omega(x^{\mu})$ must be positive for all $x^{\mu}$ we are left with only two possible solutions, the trivial $\Omega(x^{\mu})=1$ and the non-trivial $\Omega(x^{\mu})=\sqrt{1+ a/(x^{\mu})^{2}}$. Since $\Omega(x^{\mu})$ must be dimensionless, $a$ must have dimensions of length squared, and so we define a new constant $l$ with dimensions of length via $a \equiv l^{2}$. Therefore, the only non-trivial factor $\Omega(x^{\mu})$ for which the spacetime interval $ds^{2}=g_{\mu\nu} dx^{\mu}dx^{\nu}$ is invariant is

\begin{equation}\label{proof6}
\Omega(x^{\mu}) = \sqrt{1+ \left(\frac{l}{x^{\mu}}\right)^{2}}.
\end{equation}
\end{proof}

\end{subsection}

\end{section}
%%%%%%%%%%%%%%%%%%%%%%%%%%%%%%%%%%%%%%%%%%%%%%%%%%%%%%%%%%%%%%%%%%% 

%\begin{section}{When are $g_{\mu\nu}$ and $\tilde{g}_{\mu\nu}$ conformally related?}\label{condition}
\begin{section}{Connection with conformal transformations?}\label{condition}

Let's impose the condition 

\begin{equation}\label{ct1}
\tilde{g}_{\rho\sigma}=g_{\rho\sigma}\Omega^{2}(x),
\end{equation}

\noindent for some unknown function $\Omega(x)$. Coordinate transformations $x \rightarrow \tilde{x}(x)$ that satisfy Eq.~(\ref{ct1}) are defined as conformal transformations. Therefore, if 

\begin{equation}\label{ct2}
x^{\mu} \rightarrow \tilde{x}^{\mu}(x^{\mu})=x^{\mu}\sqrt{1+ \left(\frac{l}{x^{\mu}}\right)^{2}}
\end{equation}

\noindent can be shown to satisfy Eq.~(\ref{ct1}) then it is a conformal transformation. 

If coordinates transform according to Eq.~(\ref{ct2}) then we have

\begin{equation}\label{ct4}
\frac{\partial{\tilde{x}^{\rho}}}{\partial{x^{\mu}}}=\frac{\partial{\left(x^{\rho}\Omega(x^{\rho})\right)}}{\partial{x^{\mu}}}=\frac{1}{\Omega(x^{\mu})}, \qquad (\rho=\mu).
\end{equation}

\noindent Likewise,

\begin{equation}\label{ct5}
\frac{\partial{\tilde{x}^{\sigma}}}{\partial{x^{\nu}}}=\frac{\partial{\left(x^{\sigma}\Omega(x^{\sigma})\right)}}{\partial{x^{\nu}}}=\frac{1}{\Omega(x^{\nu})}, \qquad (\sigma=\nu).
\end{equation}

\noindent For $(\rho \neq \mu)$ and $(\sigma \neq \nu)$ all partial derivatives equal zero. Now, under a change of coordinates $x \rightarrow \tilde{x}(x)$ the metric tensor transforms according to

\begin{equation}\label{ct3}
\tilde{g}_{\rho\sigma}=g_{\mu\nu} \frac{\partial{x^{\mu}}}{\partial{\tilde{x}^{\rho}}} \frac{\partial{x^{\nu}}}{\partial{\tilde{x}^{\sigma}}}=g_{\mu\nu} \left(\frac{\partial{\tilde{x}^{\rho}}}{\partial{x^{\mu}}} \frac{\partial{\tilde{x}^{\sigma}}}{\partial{x^{\nu}}}\right)^{-1}.
\end{equation}

\noindent Using Eq.~(\ref{ct3}) we therefore find

\begin{equation}\label{ct6}
\tilde{g}_{\rho\sigma}=g_{\mu\nu}\Omega^{2}(x^{\mu}), \qquad (\mu=\nu).
\end{equation}

%\noindent Equation~(\ref{ct6}) therefore suggests $\tilde{g}$ and $g$ are not conformally related because different components of the metric tensor are rescaled by different functions of the coordinates, since in general $\Omega(x^{0}) \neq \Omega(x^{1}) \neq \Omega(x^{2}) \neq \Omega(x^{3})$. Therefore, the two metrics are only conformally related if $\Omega(x^{0})=\Omega(x^{1})=\Omega(x^{2})=\Omega(x^{3})$.

\noindent Although this looks very much like a conformal transformation, Eq.~(\ref{ct6}) in fact tells us that $\tilde{g}$ and $g$ are in general not conformally related, since $\Omega(x^{\mu})$ does not necessarily define the same function for each index $\mu$.

%different components of the metric tensor are rescaled by different functions of the coordinates, e.g. $\Omega(x^{0})$ is a different function than $\Omega(x^{1})$, etc. Equations~(\ref{eq1}) and~(\ref{eq2}) thus define a conformal transformation only when $\Omega(x^{\mu})$ defines the same function for each index $\mu$. 

%However, the two metrics are conformally related for the specific case of $\Omega(x^{0})=\Omega(x^{1})=\Omega(x^{2})=\Omega(x^{3})$. Equations~(\ref{eq1}) and~(\ref{eq2}) thus define a conformal transformation only in this specific case. 

We further highlight the fact that the coordinate transformation of Eq.~(\ref{ct2}) is not continuous at the origin, and so is not defined over the whole manifold as is the case for a normal conformal transformation. 

%%%%%%%%%%                                                                                                                                                                                                                                                                                                                                                                                  
\end{section}

%%%%%%%%%%%%% 

%%%%%%%%%%%%%%%%%%%%%%%%%%%%%%%%%%%%%%%

\begin{section}{Discussion and conclusions}

A conformal transformation is just a special type of coordinate transformation, a diffeomorphism~\cite{Farnsworth:2017tbz}. That is, a conformal transformation does not physically change anything at all, at least not directly, as all physical observables must be coordinate independent. So why should we care about conformal transformations? Well, conformal transformations may not directly transform physical observables, but they may have a yet greater impact by constraining the laws of physics themselves. For example, the Einstein-Hilbert action of general relativity is the simplest action that is invariant under a local change of spacetime coordinates. Symmetry and simplicity led to the mathematical formalism of general relativity. Might we play a similar game, and ask what is the simplest gravitational action that is invariant under a conformal transformation?

We already know the simplest action that is invariant under a Weyl transformation, which is similar to a conformal transformation but differs in one crucial respect. A Weyl transformation is a local rescaling of the metric tensor $g_{\mu \nu}(x) \rightarrow g_{\mu \nu}(x)\Omega^{2}(x)$, but one for which spacetime coordinates remain fixed $x \rightarrow x$~\cite{Farnsworth:2017tbz}. Thus, a Weyl transformation \emph{can} change physical observables, such as the spacetime interval via $ds^{2} \rightarrow ds^{2} \Omega^{2}(x)$. The simplest action that is invariant under a Weyl transformation is 

\begin{equation}\label{eq5}
S_{W}= -2\alpha_{G} \int \left(R^{\mu\nu}R_{\mu\nu} - \frac{1}{3}R^{2} \right)d^{4}x \sqrt{-g}, 
\end{equation}

\noindent where $\alpha_{G}$ is a dimensionless gravitational coupling and $R_{\mu\nu}$ is the Ricci tensor~\cite{Mannheim:2011ds}. This theory produces fourth-order equations for fluctuations about a fixed background, which may or may not lead to problems with unitarity~\cite{Bender:2007wu}. However, a true conformal transformation for which the four-volume element $d^{4}x \sqrt{-g}$ is invariant may require an additional multiplicative term in the integrand of Eq.~(\ref{eq5})~\cite{Dabrowski:2008kx}. This preliminary line of enquiry will be pursued in future work. 

%Since the gravitational Lagrangian should be a function of the metric tensor and its derivatives, and given that the metric tensor transforms like $g_{\mu \nu}(x^{\mu}) \rightarrow g_{\mu \nu}(x^{\mu})\Omega^{2}(x^{\mu})$, perhaps the simplest conformally invariant action is

%\begin{equation}
%S_{C}=\frac{1}{8\pi G} \int g^{2}_{\mu\nu} \left(R^{\mu\nu}R_{\mu\nu} - \frac{1}{3}R^{2} \right)d^{4}x \sqrt{|g_{\mu \nu}|}.
%\end{equation} 

%\noindent This preliminary line of enquiry will be pursued in future publications.

Conformal symmetry is already present in many physical theories, either as an exact or approximate symmetry. For example, the standard model of particle physics exhibits an underlying conformal symmetry that is only broken by the presence of massive scalar fields at lower energies. Maxwell's equations are conformally invariant in $4$-dimensional spacetime, the ultra-relativistic limit of special relativity exhibits an effective conformal invariance, and string theory includes conformal symmetry at a fundamental level via the AdS/CFT correspondence. An important and attractive feature of conformal transformations is that the angle between vectors is scale invariant. Therefore, conformal transformations leave the light-cone structure unchanged, ensuring the preservation of causality and an invariant speed of light at all distances scales (see Refs.~\cite{Coumbe:2015zqa,Coumbe:2015aev,Coumbe:2015bka,Coumbe:2018myj} for more details). Given its high degree of symmetry and ubiquity, conformal symmetry may prove to be a fruitful guide in the future development of fundamental physics.

%This is useful in that it allows one to compute the geodesic motion of light-like particles via the ultrarelativistic limit of time-like massive particles.

However, formulating a conformally symmetric theory comes with a wide set of challenges that are beyond the scope of the present work, but are nevertheless important enough to mention. Firstly, general relativity contains a fixed length scale associated with the gravitational coupling $G$, which may be thought of as defining an absolute background structure. Conformal invariance seems to require removing all such length scales in order to achieve true scale invariance. Another difficulty is that conformal invariance can only ever be an exact symmetry in the high-energy limit, since it must be broken at lower energies due to the existence of particles of non-zero mass that can be associated with a length scale. A well-defined symmetry breaking mechanism is therefore required. Lastly, there is the so-called conformal anomaly problem~\cite{Deser:1996na}. 

In this work, the only non-trivial conformal-like transformation, as defined via Eqs.~(\ref{eq1}) and~(\ref{eq2}), consistent with an invariant line element is determined.

\end{section}

%%%%%%%%%%%%%%%%%%%%%%%%%%%%%%%%%%%%%%%%%%%%%%%%%%%%%%%%%%%%%%%%%%%                                                                                                               
%%%%%%%%%%%%%%%%%%%%%%%%%%%%%%%%%%%%%%%%%%%%%%%%%%%%%%%%%%%%%%%%%%%                                                                                                    
%%%%%%%%%%%%%%%%%%%%%%%%%%%%%%%%%%%%%%%%%%%%%%%%%%%%%%%%%%%%%%%%%%%

\section*{Acknowledgements}

I wish to thank Roberto Percacci for his useful comments. I acknowledge support from the Danish Research Council grant “Quantum Geometry”. Thanks to all participants of the Fifth International Conference on the Nature and Ontology of Spacetime.

%%%%%%%%%%%%%%%%%%%%%%%%%%%%%%%%%%%%%%%%%%%%%%%%%%%%%%%%%%%%%%%%%%%                                                                                                                         %%%%%%%%%%%%%%%%%%%%%%%%%%%%%%%%%%%%%%%%%%%%%%%%%%%%%%%%%%%%%%%%%%%                                                                                                                         %%%%%%%%%%%%%%%%%%%%%%%%%%%%%%%%%%%%%%%%%%%%%%%%%%%%%%%%%%%%%%%%%%%

\bibliographystyle{unsrt}
\bibliography{Master}

\begin{thebibliography}{10}

\bibitem{Gross123}
D.~J. {Gross}.
\newblock {The Role of Symmetry in Fundamental Physics}.
\newblock {\em Proceedings of the National Academy of Science},
  93:14256--14259, December 1996.

\bibitem{PhysRev.159.1251}
Sidney Coleman and Jeffrey Mandula.
\newblock All possible symmetries of the $s$ matrix.
\newblock {\em Phys. Rev.}, 159:1251--1256, Jul 1967.

\bibitem{Dabrowski:2008kx}
Mariusz~P. Dabrowski, Janusz Garecki, and David~B. Blaschke.
\newblock {Conformal transformations and conformal invariance in gravitation}.
\newblock {\em Annalen Phys.}, 18:13--32, 2009.

\bibitem{Farnsworth:2017tbz}
Kara Farnsworth, Markus~A. Luty, and Valentina Prilepina.
\newblock {Weyl versus Conformal Invariance in Quantum Field Theory}.
\newblock {\em JHEP}, 10:170, 2017.

\bibitem{Mannheim:2011ds}
Philip~D. Mannheim.
\newblock {Making the Case for Conformal Gravity}.
\newblock {\em Found. Phys.}, 42:388--420, 2012.

\bibitem{Bender:2007wu}
Carl~M. Bender and Philip~D. Mannheim.
\newblock {No-ghost theorem for the fourth-order derivative Pais-Uhlenbeck
  oscillator model}.
\newblock {\em Phys. Rev. Lett.}, 100:110402, 2008.

\bibitem{Coumbe:2015zqa}
D.~N. Coumbe.
\newblock {Hypothesis on the Nature of Time}.
\newblock {\em Phys. Rev.}, D91(12):124040, 2015.

\bibitem{Coumbe:2015aev}
Daniel Coumbe.
\newblock {Quantum gravity without vacuum dispersion}.
\newblock {\em Int. J. Mod. Phys.}, D26(10):1750119, 2017.

\bibitem{Coumbe:2015bka}
Daniel Coumbe.
\newblock {What is dimensional reduction really telling us?}
\newblock In {\em {14th Marcel Grossmann Meeting on Recent Developments in
  Theoretical and Experimental General Relativity, Astrophysics, and
  Relativistic Field Theories (MG14) Rome, Italy, July 12-18, 2015}}, 2015.

\bibitem{Coumbe:2018myj}
D.~N. Coumbe.
\newblock {Renormalizing Spacetime}.
\newblock {\em Int. J. Mod. Phys.}, D28:1950008, 2019.

\bibitem{Deser:1996na}
Stanley Deser.
\newblock {Conformal anomalies: Recent progress}.
\newblock {\em Helv. Phys. Acta}, 69(4):570--581, 1996.

\end{thebibliography}

%\end{multicols}

\end{document}